\documentclass{jps-cp}
\graphicspath{{logos/},{figs/}}
\title{Design and Performance of HV CMOS Sensors for Future Colliders by the RD50 Collaboration}

\author{Jory \textsc{Sonneveld}$^{1}$ for the RD50 Collaboration}

\inst{$^{1}$Nikhef National Institute for Subatomic Physics, Science Park 105, 1098 XG Amsterdam, the Netherlands}

\email{jory.sonneveld@nikhef.nl}

\recdate{January 20, 2023}

\abst{

The CERN RD50 collaboration develops depleted monolithic active pixel CMOS sensors for future colliders with the aim of high radiation tolerance, good time resolution, and high granularity pixel detectors. The most recent prototype, the RD50-MPW3, is a 150 nm High Voltage CMOS LFoundry chip that features pixels with a 62 $\mu$m pitch that integrate both digital and analog readout electronics inside the sensing diodes. The 64 x 64 pixels on this chip are arranged in 32 double columns and have an optimized periphery for efficient configuration and fast serial data transmission. Post-layout simulations of a single pixel show a power consumption of 22 $\mu$W per pixel and 9 ns time walk.

The predecessor of this version, the RD50-MPW2, was shown to match simulation results in tests at beam facilities and to have a time resolution of 300 ps both before and after irradiation to a fluence of  $\Phi_{\mathrm{eq}} = 5\cdot     10^{14}/\mathrm{cm}^2$. This proceeding discusses the design of the latest advanced prototype, the RD50-MPW3, the first results for the RD50-MPW3, and the performance of the RD50-MPW2.
}

\kword{CMOS, DMAPS, sensors, pixel detectors, semiconductors, radiation hardness, timing detectors}

\begin{document}
\maketitle

\section{Introduction}
The RD50 collaboration  develops radiation hard semiconductor devices for very high luminosity colliders\cite{rd50}. Detectors will face 1 MeV neutron equivalent fluences of $\Phi_{\mathrm{eq}} > 10^{16}/\mathrm{cm}^2$ at the future High Luminosity Large Hadron Collider (HL-LHC) \cite{hllhc} and up to $\Phi_{\mathrm{eq}} = 6\cdot 10^{17}/\mathrm{cm}^2$ at the Future Circular hadron Collider (FCC-hh) \cite{ecfa,fccsnowmass}. In this light, RD50 develops sensors that can withstand these fluences. One of the areas that the RD50 collaboration focuses on is new structures such as High Voltage CMOS (HV CMOS). The CERN RD50 HV CMOS group aims to further improve and demonstrate the performance of these sensors with their work on ASIC design, TCAD simulations, DAQ development, and chip performance evaluation.

\begin{figure}[tbh]
    \centering
    \includegraphics[width=0.47\textwidth]{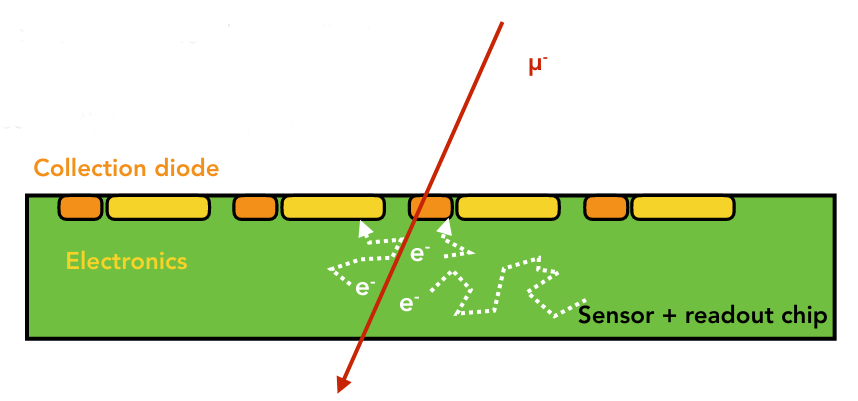}
    \includegraphics[width=0.47\textwidth]{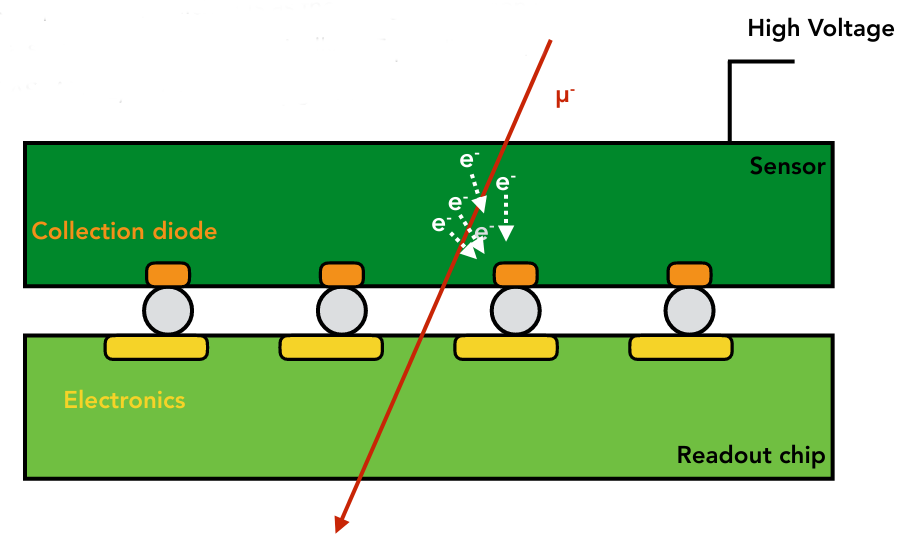}
    \caption{Left: A monolithic active pixel sensor, where electronics is integrated into the sensor. No processing is needed and this technology is commercially available. Right: a hybrid pixel sensor, where a silicon sensor is bump bonded to an ASIC. This leaves more space for per pixel functionalities. Figures from \cite{hynds}.}
    \label{fig:monolithichybrid}
\end{figure}

Monolithic Active Pixel Sensors, or MAPS for short, have electronics integrated into a sensor resulting in a low mass detector that does not require any processing such as bump bonding, as shown on the left in Fig. \ref{fig:monolithichybrid}. It is widely used in digital cameras like those found in smartphones, making the commercial process accessible at low-cost. This is in contrast to hybrid pixel detectors, where a sensor is bump bonded to an Application Specific Integrated Circuit (ASIC), as shown on the right in Fig. \ref{fig:monolithichybrid}. Such a sensor, however, has more room for per pixel functionalities on the separate ASIC because of the lack of HV rings around each readout cell.

\subsection{DMAPS}
\begin{figure}[tbh]
    \centering
    \includegraphics[width=0.47\textwidth]{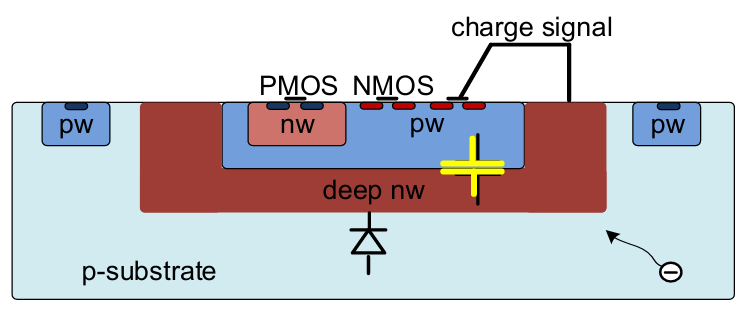}
    \includegraphics[width=0.47\textwidth]{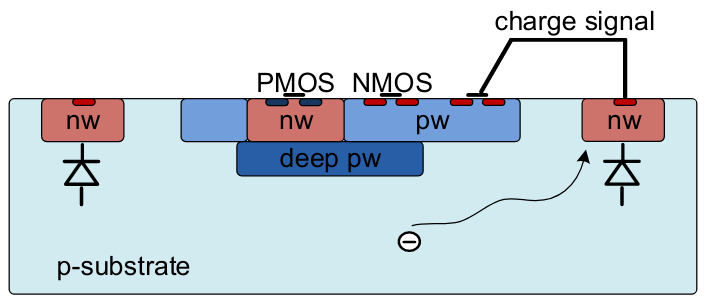}
    \caption{Left: A depleted monolithic sensor with a large collection electrode, resulting in a strong electric field and homogeneous weighting field. Right: A DMAPS with a small collection electrode, resulting in low noise and low capacitance. Figure from \cite{monopix}.}
    \label{fig:coll}
\end{figure}

Depleted Monolithic Active Pixel Sensors, or DMAPS for short \cite{dmaps}, as well as High Resistivity CMOS (HR-CMOS), have a bias voltage applied to deplete the sensor and create a region sensitive to particle signals. Two distinct developments make use of this feature. One, using high voltage in processes developed for power electronics and with a large fill factor \cite{evavertex18}, has a large deep n-well in a high resistivity p-substrate, see on the left in Fig. \ref{fig:coll}. This type of DMAPS is also sometimes called HV CMOS after the process used. The deep n-well is for charge collection, and the substrate can be biased up to hundreds of volts. The p-n junction is at the interface between the large deep n-well and the p-substrate. The readout electronics is placed inside the charge collection well.

The other development uses small charge fill factor DMAPS with substrates and thick epitaxial layers from CMOS imaging processes, as shown on the right in Fig. \ref{fig:coll}. Small n-wells are placed inside p-substrate that can be biased to several volts. The p-n junction is at the interface between the n-well and the p-substrate. The charge signal is routed from the small n-well to the readout electronics, which is separate from the small n-well. This design results in less cross talk than when the electronics is inside the collection electrode, and the small size of the n-well also leads to a very small sensor capacitance of about 5 fF only, leading to low noise and low power use.

After certain fluences of particles, non-ionizing radiation damages the silicon sensors and defects can result in the trapping of charge signals \cite{trapping}. A stronger electric field results in less trapping of charges and thus a higher radiation tolerance. A large electric field, however, comes at the cost of more capacitance, power, and noise: a large collection well results in a sensor capacitance of more than 100 fF.

\subsection{RD50 HV CMOS}
\begin{figure}[tbh]
    \centering
    \includegraphics[width=0.57\textwidth]{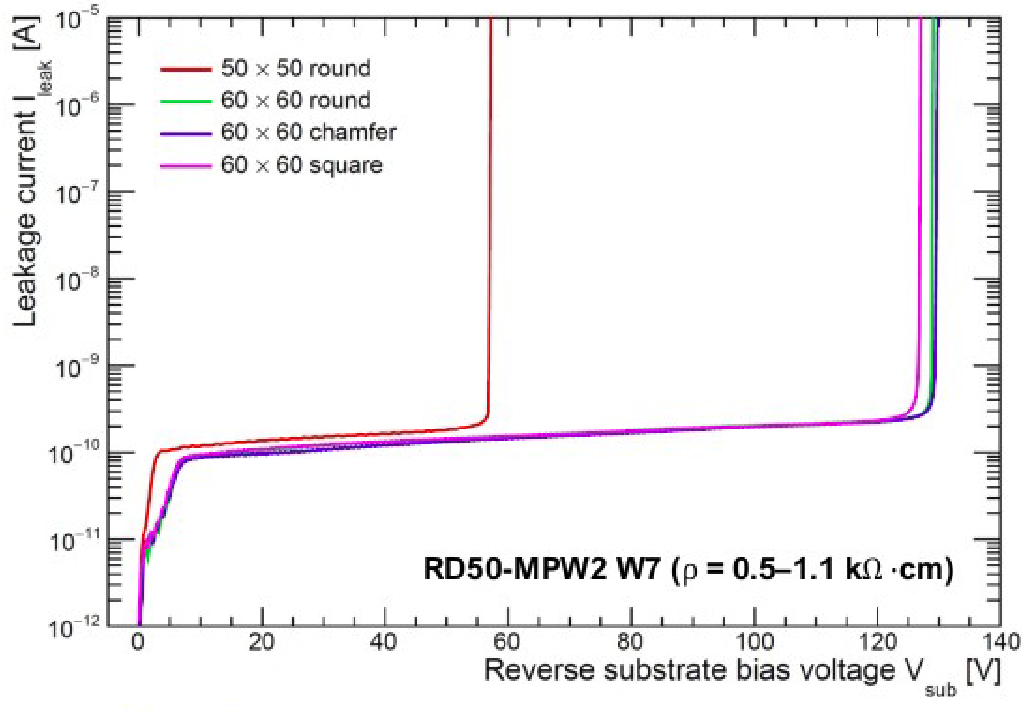}
    \caption{Three current-voltage (IV) curves for the second submission of the RD50 DMAPS chip. More spacing between electrodes resulting in larger pixels and rounded pixel corners improved the breakdown voltage of the sensor. Figure from \cite{hernandezvertex2020}.}
    \label{fig:breakdownoptimization}
\end{figure}

The RD50 collaboration develops HV CMOS sensors. A large fill factor CMOS sensor was chosen for fast collection time, high radiation tolerance, and high breakdown voltage. Currently, there are many developments on HV CMOS in high energy physics. These efforts have achieved sensors that are radiation tolerant to fluence of $\Phi_{\mathrm{eq}} = 10^{15}/\mathrm{cm}^2$, precise in time to 100 picoseconds \cite{cactus} or even tens of picoseconds for a high voltage, high resistivity SiGe CMOS sensor with gain \cite{monolith}, and a breakdown voltage of no less than 400 V \cite{lfmonopix}. The RD50 collaboration has improved its HV CMOS design to optimize the breakdown voltage by adding rounded pixel corners and increasing the spacing between electrodes to 8 $\mu$m up from 3$~\mu$m in the RD50-MPW1. The effect can be seen in Fig. \ref{fig:breakdownoptimization}.

\begin{figure}[tbh]
    \centering
    \includegraphics[width=0.63\textwidth, trim=7 5 17 28, clip]{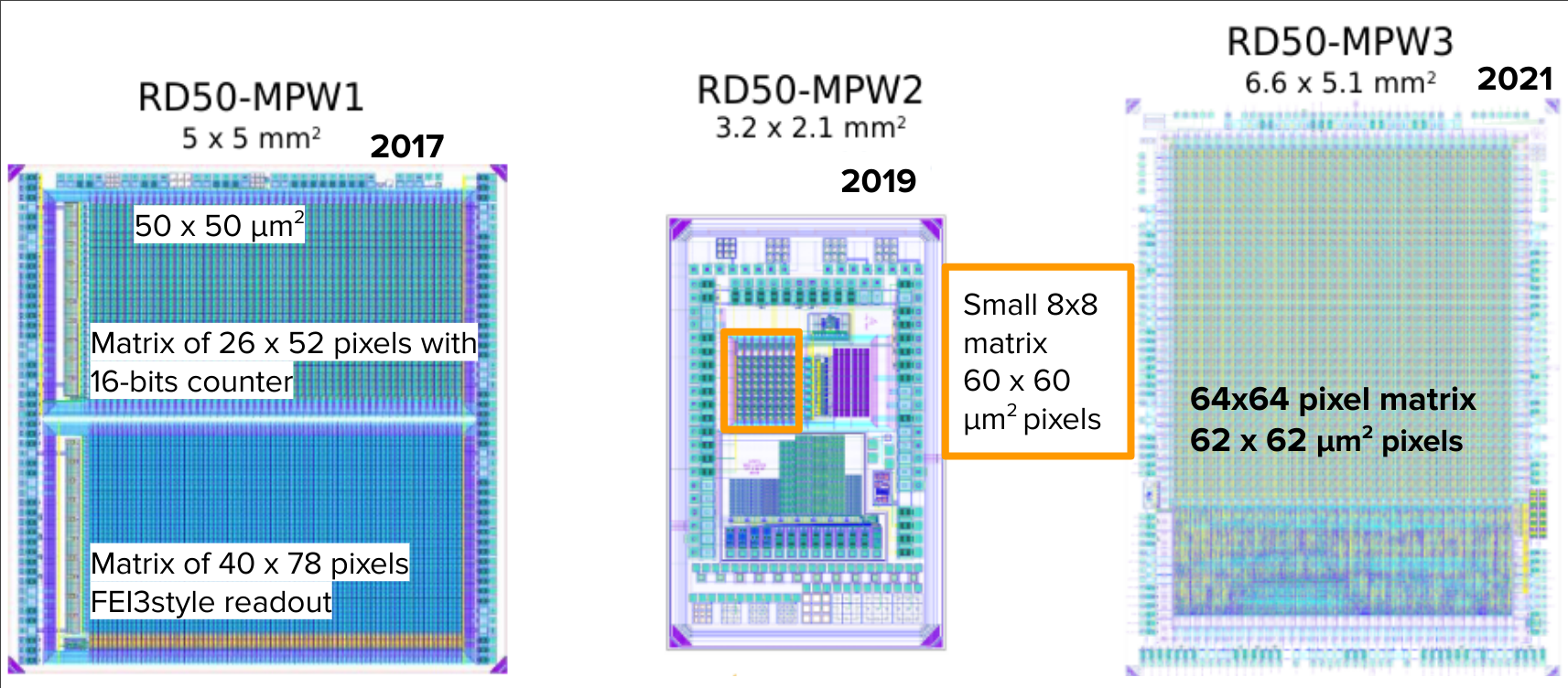}
\caption{Three submissions of the RD50 HV CMOS monolithic sensors. The latest has an in-pixel digital readout and an advanced digital periphery. Figure from \cite{debevcrd50}.}
    \label{fig:submissions}
\end{figure}

There were three submissions of the RD50 HV CMOS monolithic sensors, as shown in Fig. \ref{fig:submissions} and listed in Table \ref{tab:submissions}. All are fabricated in the LF15A process from LFoundry S.r.l. in 150 nm HV CMOS. In 2017, a $5\times5$ mm$^2$ chip was submitted with two pixel matrices: one $26\times52$ pixel matrix with $75~\mu\mathrm{m}\times75~\mu$m pixels and a 16 bits counter and a $40 \times 78$ pixel matrix with $50~\mu\mathrm{m}\times50~\mu$m pixels and an FEI3-style readout. It had, however, a low breakdown voltage of only 60 V, and a high leakage current of the order of $\mu$A. In 2019, a $3.2\times2.1$ mm$^2$ chip with an $8 \times 8$ pixel matrix with 60 $\mu$m pixel pitch was submitted. This had a significantly higher breakdown voltage of 120 V, and a leakage current of only of the order of nA per pixel. It had an analog readout only.

\begin{table}[h!]
    \begin{center}
        \caption{Summary of the three RD50 MAPS chip submissions in 150 nm HV CMOS.}
        \label{tab:submissions}
        \begin{tabular}{c|c|c|c|c } 
            \textbf{Chip name} & \textbf{Submission year}  & \textbf{Pixel pitch} & \textbf{Pixel matrix size}  & \textbf{Chip size} \\ \hline \\ 
        RD50-MPW1 & 2017 & 50 $\mu$m, 75 $\mu$m & $26\times52$, $40\times78$ pixels  & $5\times5$ mm$^2$     \\ 
        RD50-MPW2 & 2019 & 60 $\mu$m & $8\times8$ pixels    & $3.2\times2.1$ mm$^2$ \\ 
        RD50-MPW3 & 2021 & 62 $\mu$m & $64\times64$ pixels  & $6.6\times5.1$ mm$^2$ \\ 
        \end{tabular}
    \end{center}
\end{table}

\section{The latest RD50 HV CMOS design}
\begin{figure}[tbh]
    \centering
    \includegraphics[width=0.22\textwidth]{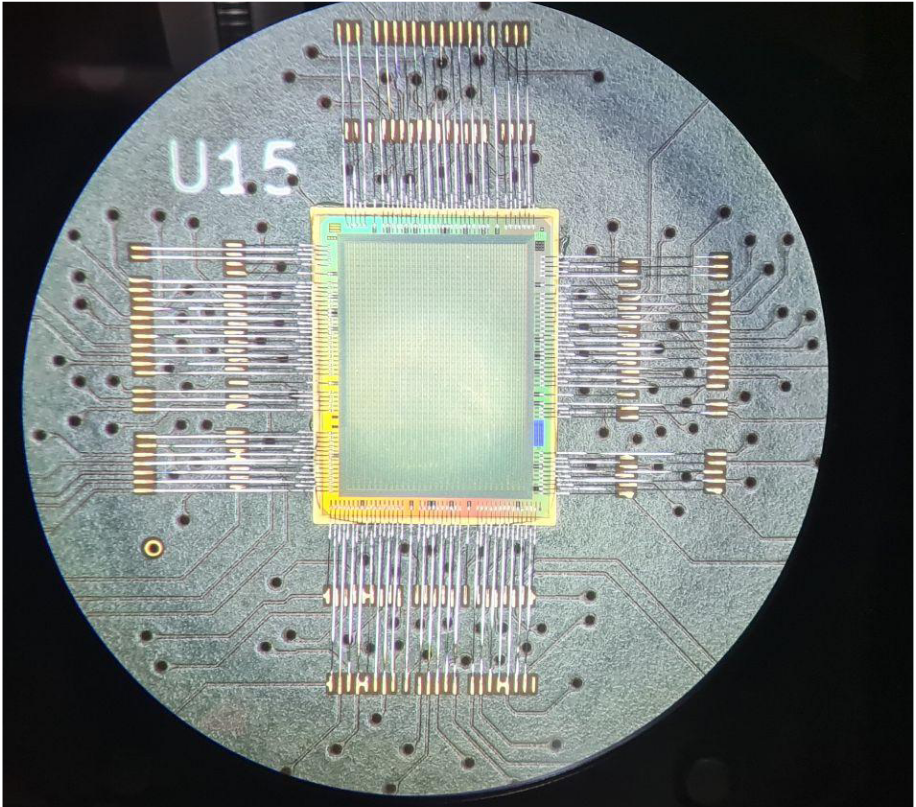}
    \includegraphics[width=0.52\textwidth]{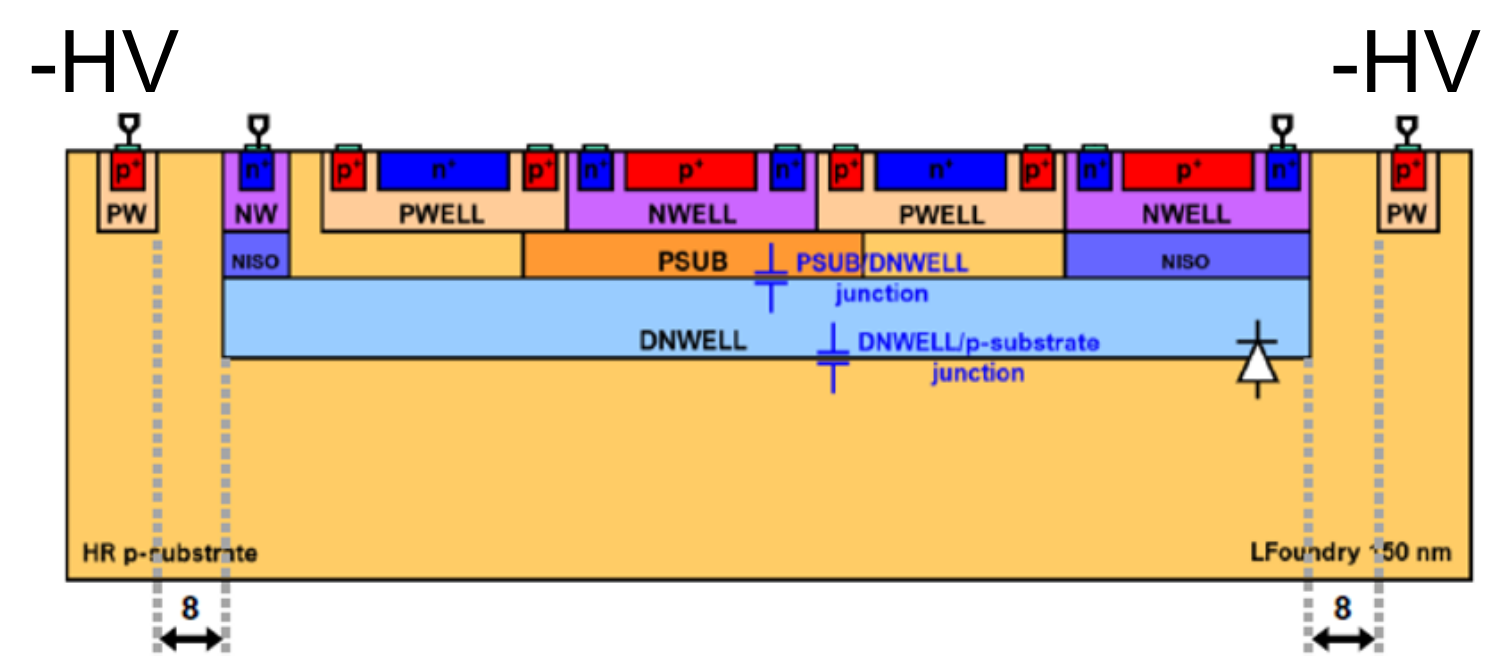}
    \caption{The latest HV CMOS submission of the RD50 collaboration. Left: Picture of the RD50-MPW3. Courtesy of Uwe Kr\"{a}mer. Right: Schematic cross section of the sensor from \cite{chenfanrd50}.}
    \label{fig:mpw3}
\end{figure}

More recently in 2021, a third design, called the Multi-Project Wafer 3 or RD50-MPW3 for short, was submitted and was recently analyzed in the laboratory. A picture is shown in Fig \ref{fig:mpw3}.  This $6.6\times 5.1$ mm$^2$ structure has a $64 \times 64$ matrix with 62 $\mu$m pitch pixels with digital readout, and an advanced peripheral readout\cite{evavertex21,siebererzhangtwepp22}. The slightly increased pixel pitch allows for the extra features while still keeping the electrode spacing 8 $\mu$m. The RD50-MPW3 was produced in one standard and two high resistivities of 10 $\Omega\cdot$cm, 1.9 k$\Omega\cdot$cm and 3 k$\Omega\cdot$cm. High voltage is applied from the top, and the analog and digital electronics are inside the collection electrode that covers 55\% of the pixel area; see also Fig. \ref{fig:mpw3}. From post-layout simulations, the capacitance is 250 fF and the simulated power use is 22~$\mu$W per pixel with a digital voltage supplied at 1.8 V. The simulated gain is 230 mV, and the simulated time over threshold is 55 ns, both for a 5 ke$^-$ signal. The simulated noise is 120 e$^-$, and the simulated timewalk is 9 ns.

\subsection{Minimization of crosstalk and voltage drop}
\begin{figure}[tbh]
    \centering
    \includegraphics[width=0.52\textwidth]{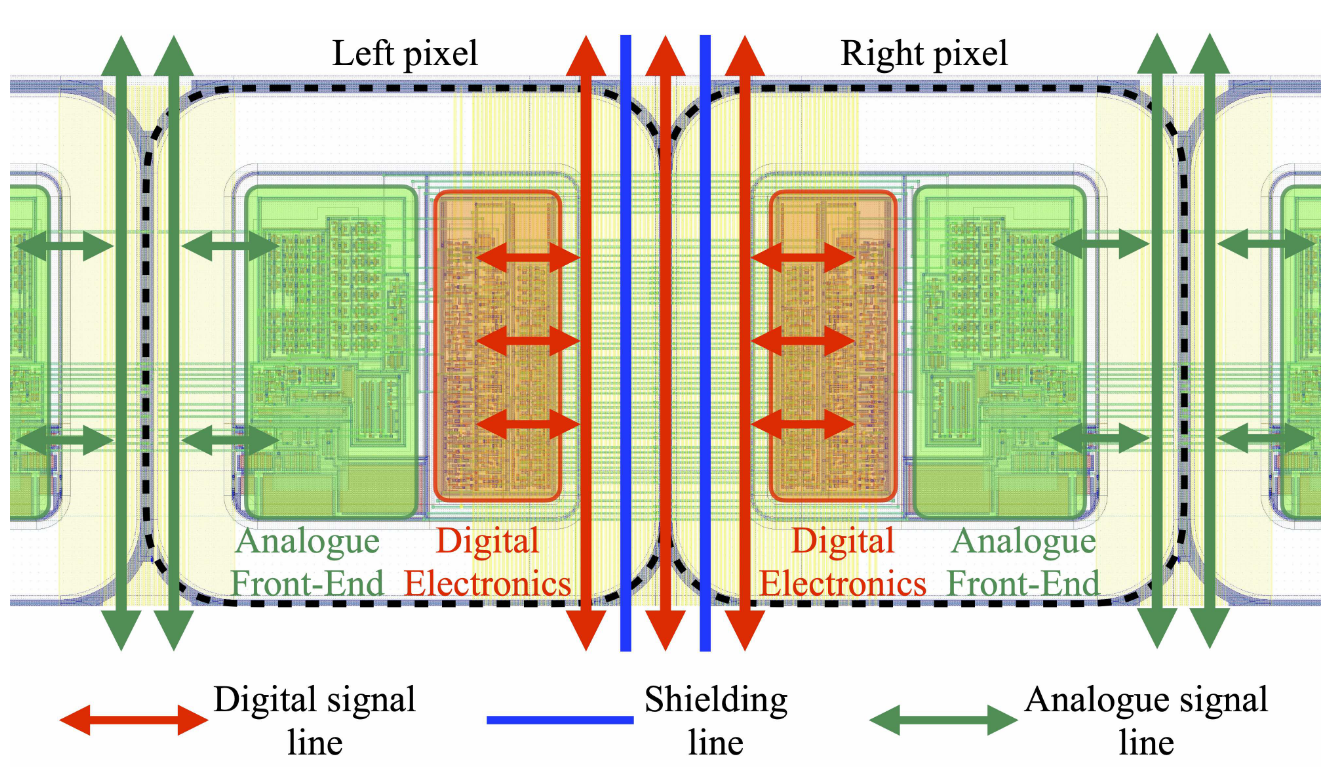}
    \caption{The analog and digital signal lines are routed along different paths in pixel double columns to prevent crosstalk. Digital lines (red) are routed in the middle of a double column and separated by grounded shielding lines. Analog signal lines (green) are shared between two double columns. Two pixels of a double column are shown here. Figure from \cite{siebererzhangtwepp22}.}
    \label{fig:analogdigital}
\end{figure}

The latest RD50 HV CMOS development has analog and digital circuits in deep p-wells separated by a trench made of shallow n-well and deep n-isolation (called NISO in this process), with their signal lines well-separated to minimize crosstalk. The RD50-MPW3 has a double column drain architecture with 128 pixels in one double column. Two pixels in a double column are shown in Fig. \ref{fig:analogdigital}. The digital lines are routed through the middle of a double column and separated by grounded shielding lines, whereas the analog signal lines are shared between two double columns but used by two pixels only. Additionally, after a voltage drop of 200 mV was observed along a power rail in the RD50-MPW1, the RD50-MPW3 features a power grid for an even power distribution with power from a ring supplied from all sides to minimize the voltage drop.

\subsection{Front-end design} 
\begin{figure}[tbh]
    \centering
    \includegraphics[width=\textwidth, trim=20 33 24 43, clip]{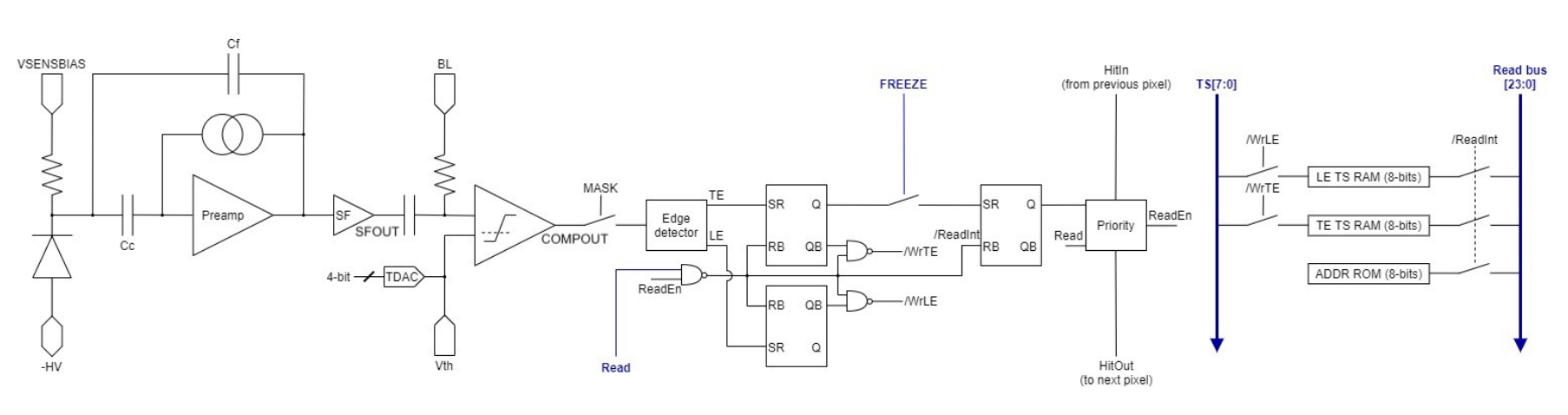}
    \caption{Schematic of the in-pixel front-end of the RD50-MPW3 chip. The analog front-end is shown on the left and the digital front-end on the right. In contrast to the previous prototype RD50-MPW2 that features continuously and switched reset pixels (see Fig. \ref{fig:picturetot}), this chip has a continuously reset charge sensitive amplifier only. Figure from \cite{siebererzhangtwepp22}.}
    \label{fig:mpw3_frontend}
\end{figure}

A schematic of the pixel front-end is shown in Fig. \ref{fig:mpw3_frontend}. Each pixel in the RD50-MPW3 has a continuous reset charge sensitive amplifier with a high processing speed and a comparator to digitize the signal. A 4-bit trim digital to analog converter (DAC) can be used to adjust individual pixel thresholds, and an injection circuit allows for performance characterization.

A pixel can be masked, as well as frozen to pause processing of new hits until a previous event in a double column is read out. The time of arrival and time over threshold are recorded in each pixel, and two 8-bit timestamps for the leading and trailing edge, respectively in 8-bit RAMs are sent out with an 8-bit pixel address via a shared 24-bit readout bus. This results in an in-pixel digital readout, that is then sent out with a double column drain architecture and a rolling shutter.

\subsection{Digital periphery and readout} Each double column has its own End Of Column (EOC) that handles readout and configuration of that column. The EOC has a FIFO memory that allows for immediate pixel readout as long as this FIFO is not full. The triggerless readout can be in so-called normal mode, with the higher pixel address prioritized, or ``freeze'' mode, where pixel readout is frozen until the double column is read out. A readout serializer in the advanced readout periphery sends out the data at a maximum rate of 640 Mbit/s. The control and status registers of the periphery are read out and configured through a Wishbone bus using an I2C interface.
The RD50-MPW3 uses a Caribou board for the data acquisition \cite{caribou} that is connected to a Zilinx ZC706 or ZC702 evaluation board with an FPGA Mezzanine Card (FMC) on one side and a chip board on the other. The RD50-MPW3 carrier board can be extended with a second board on top. With a PC, the RD50-MPW3 can be configured and its data received over an Ethernet cable.

\subsection{Test structures} 
\begin{figure}[tbh]
    \centering
    \includegraphics[width=0.27\textwidth, trim=2 0 4 3, clip]{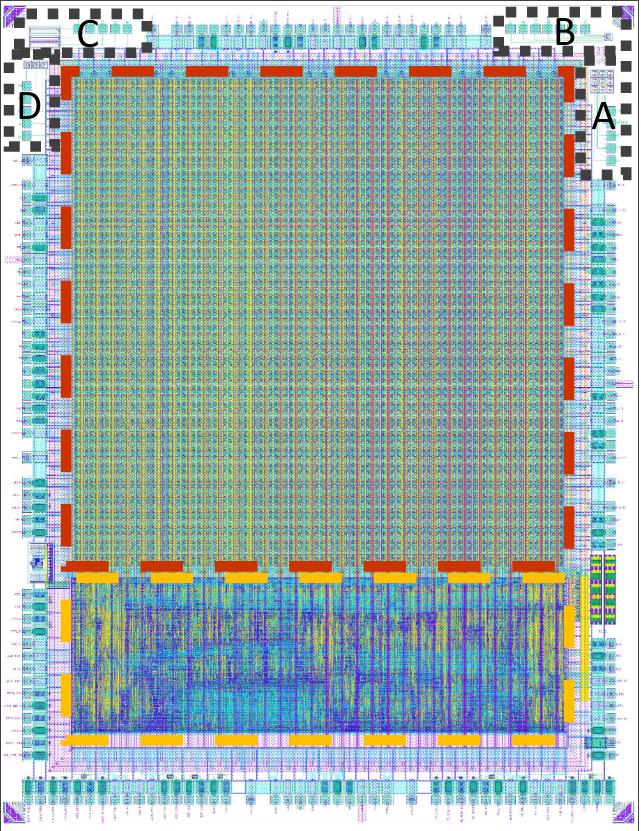}
    \caption{The RD50-MPW3 has different test structures for characterization of A: timing and charge using edge-TCT, B: capacitance using on-chip readout electronics, and C+D: defects using thermally stimulated current. Figure from \cite{siebererzhangtwepp22poster}.}
    \label{fig:mpw3test}
\end{figure}

The RD50-MPW3 features test structures for characterization, as shown in Fig. \ref{fig:mpw3test}. It contains (A) a pad for measurements of timing and charge using the edge transient current technique (eTCT). It also has a single sensing diode for measuring the capacitance of a single pixel (B) using on-chip readout electronics. Finally, it has two test structures (C+D) for defect characterization using thermally stimulated current measurements.

\subsection{First results} 
\begin{figure}[tbh]
    \centering
    \includegraphics[width=0.47\textwidth]{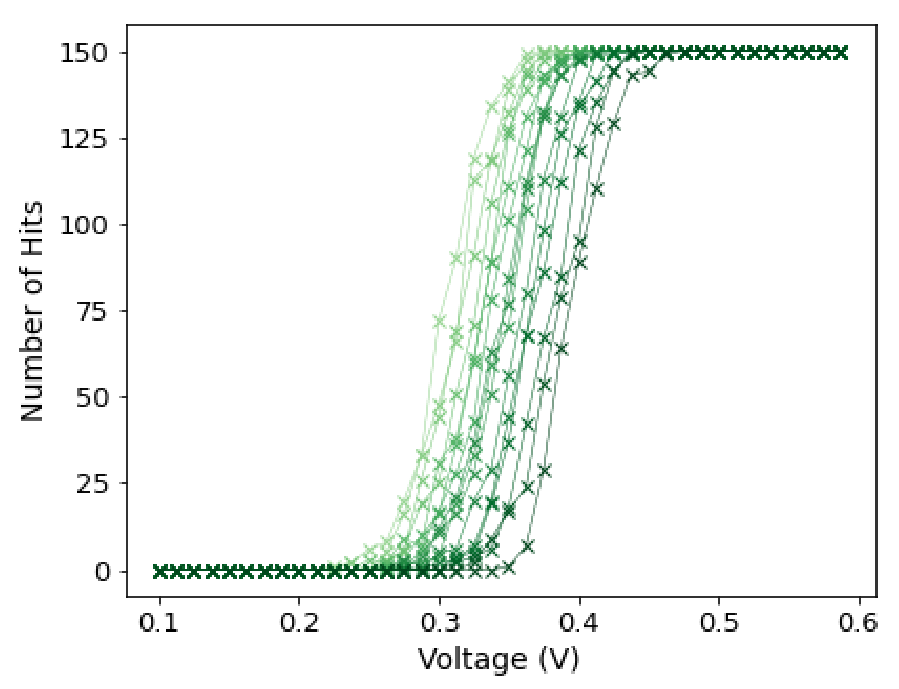}
    \includegraphics[width=0.47\textwidth, trim=3 0 0 3, clip]{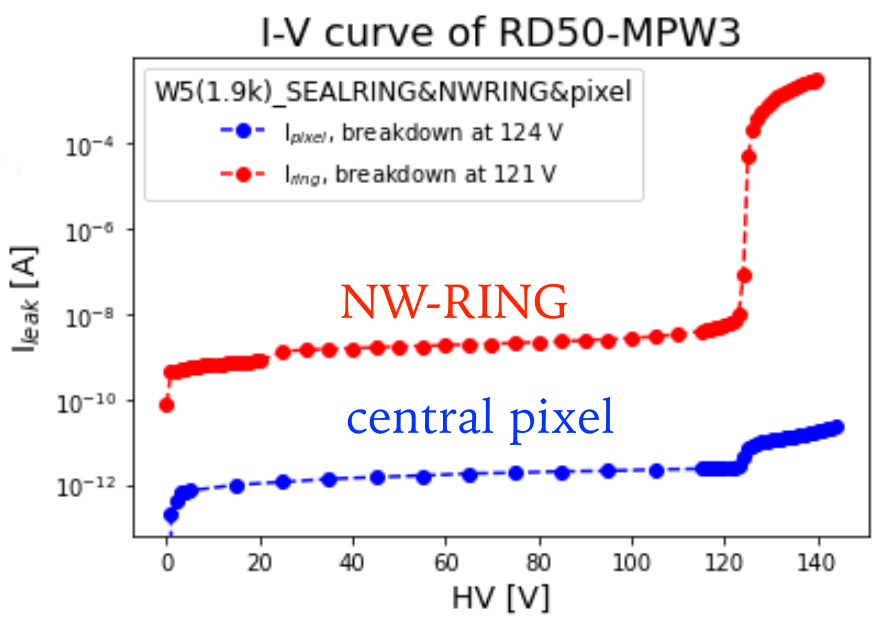}
    \caption{The first results of the RD50-MPW3. Left: The threshold for one pixel is varied with the trim DAC, where higher set thresholds are darker and 150 hits are taken for each point. As expected, the threshold increases for a higher set threshold using the trim DAC. Figure from \cite{siebererzhangtwepp22}. Right: IV curve for the n-well ring (red) and for one pixel (blue). The n-well routes away most of the leakage current and the breakdown voltage is around 120 V. Figure from \cite{workshop41rd50}.}
    \label{fig:scurveiv}
\end{figure}

The first results show that the pixel thresholds behave as expected for a non-irradiated sensor, see Fig. \ref{fig:scurveiv}. The breakdown voltage of the sensor is around 120 V with a per pixel current of only 1 pA. An n-well ring for routing away most of the pixel leakage current is effective in reducing the leakage current by four orders of magnitude, as shown in Fig. \ref{fig:scurveiv}. The RD50-MPW3 was also measured in a charged particle beam at CERN SPS with 120 GeV hadrons. Multichannel readout allows for tracking and efficiency measurements for this large pixel matrix prototype. The RD50-MPW3 was measured using an AIDA telescope with an AIDA TLU trigger and an FEI4 reference plane in EUDAQ2 \cite{eudaq2}. Data is now being analyzed and results will be published in a separate paper.

\section{Performance of the previous RD50 HV CMOS submission}
\begin{figure}[tbh]
    \centering
    \includegraphics[width=0.34\textwidth]{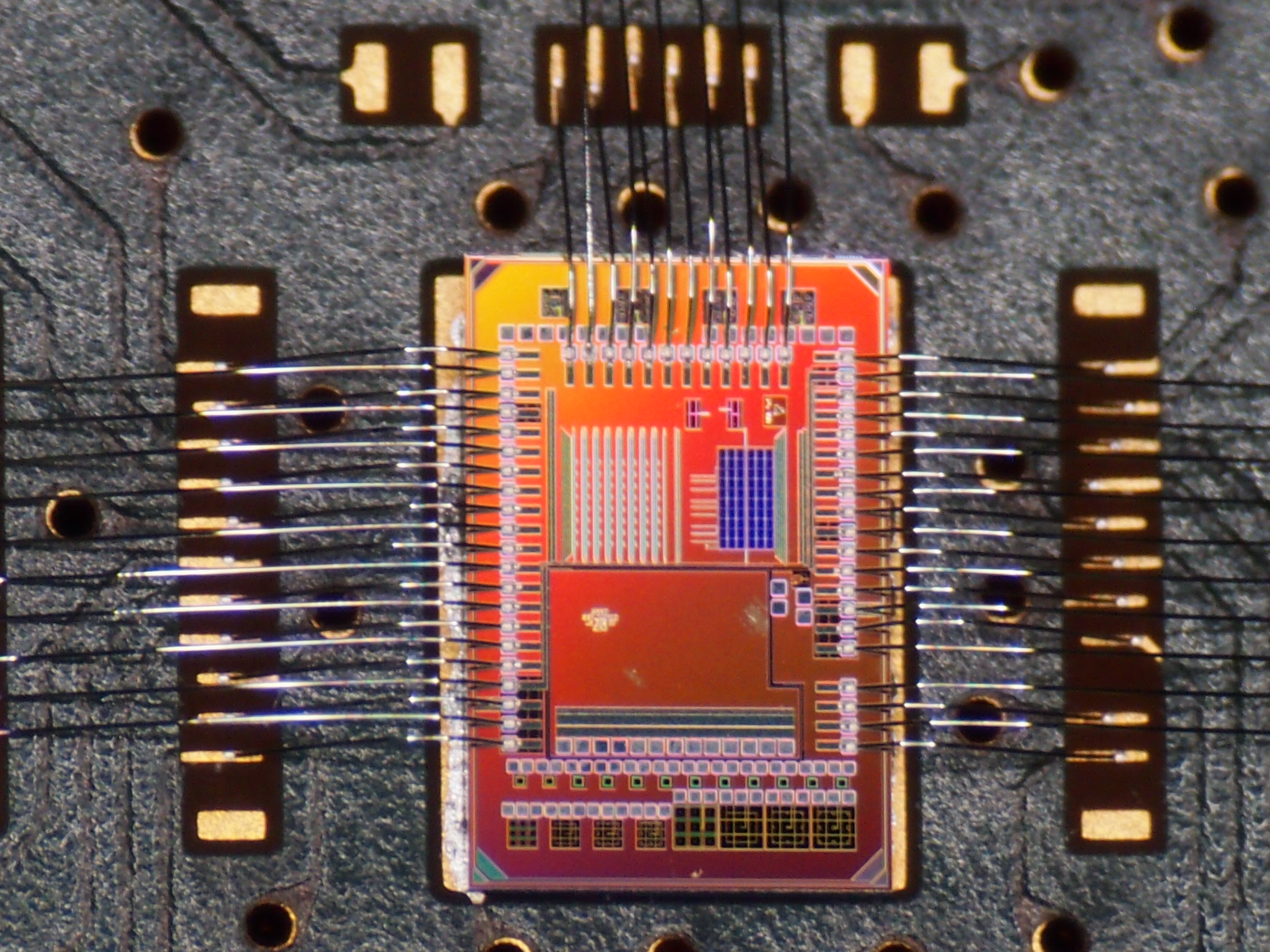}
    \hspace{2cm}
    \includegraphics[width=0.37\textwidth]{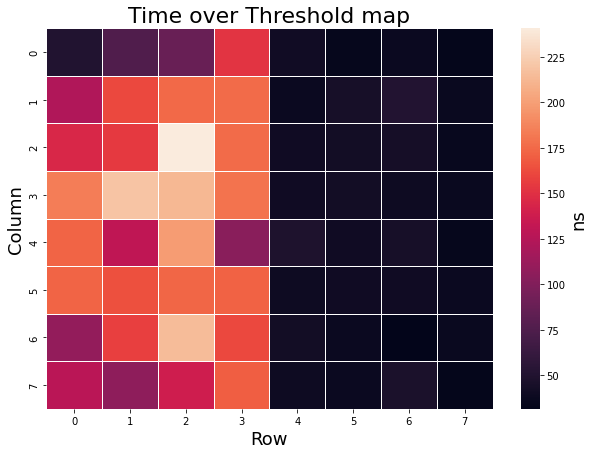}
    \caption{Left: picture of the RD50-MPW2. Picture courtesy of Christina Tsolanta. Right: time over threshold (ToT) map on all 8x8 pixels. The right 4 columns are switched reset pixels where ToT does not scale with injected charge.}
    \label{fig:picturetot}
\end{figure}

The previous submission of the RD50 HV CMOS group, also called the Multi-Project Wafer 2 or RD50-MPW2 for short, shown in Fig. \ref{fig:picturetot}, had two different types of pixels with a switched and a continuous reset, respectively. The resulting time over threshold map for all pixels can be seen in Fig. \ref{fig:picturetot}. The continuous reset pixel time over threshold scales with the signal size, and the switched reset pixel has a much faster reset.

\begin{figure}[tbh]
    \centering
    \includegraphics[width=0.97\textwidth, trim= 13 0 0 0, clip]{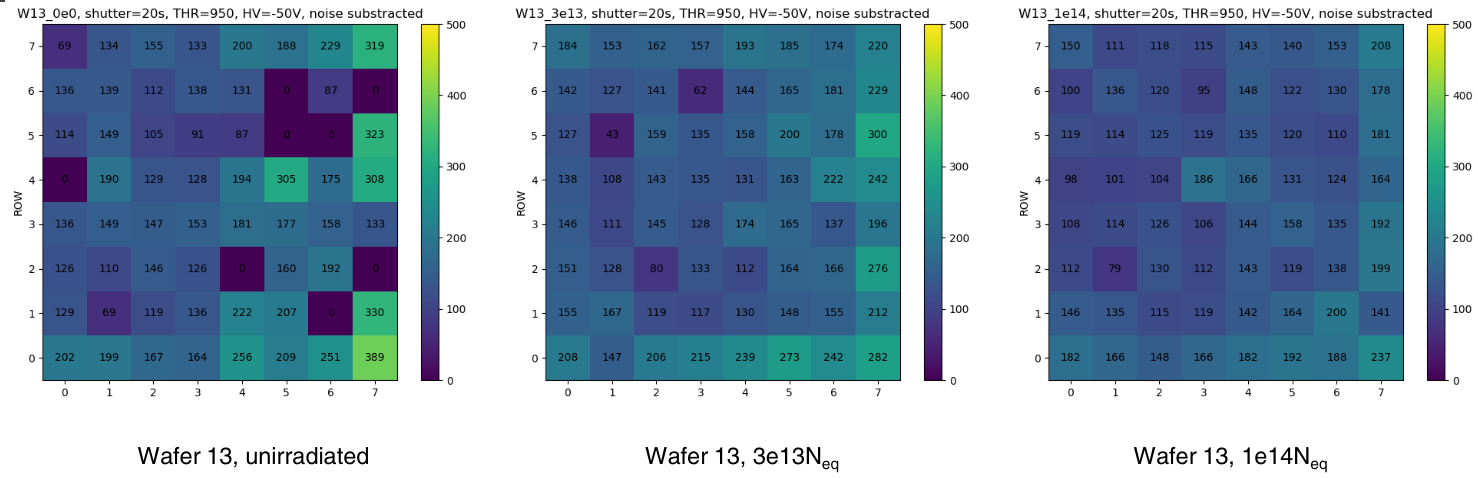}
    \caption{Noise-subtracted hit maps of the RD50-MPW2 at -50 V bias voltage measured with an $^{90}$Sr source of 370 MBq before and after irradiation to fluences $\Phi_{\mathrm{eq}} = 3\cdot10^{13}/\mathrm{cm}^2$ and  $\Phi_{\mathrm{eq}} = 10^{14}/\mathrm{cm}^2$, respectively. The pixels still respond after irradiation. Figure from \cite{38rd50}.}
    \label{fig:strontium}
\end{figure}

The RD50-MPW2 was measured with a $^{90}$Sr source after irradiation up to a fluence of  $\Phi_{\mathrm{eq}} = 10^{14}/\mathrm{cm}^2$. Although the performance shows some degradation, it still measured signals at different irradiation levels, as shown in the hit maps in Fig. \ref{fig:strontium} that were taken at -50 V bias using a strontium source.
The RD50-MPW2 was also measured in a beam test at 252.7 MeV proton beam at MedAustron with a 30 kHz event rate. The ToT measurement peaks at 35 ns \cite{sieberer22} and matches simulation \cite{siebereretal22}. As the readout of the RD50-MPW2 is only possible for 1 pixel at a time, measurements relying on tracking were not further pursued.

The RD50-MPW2 also showed only small changes in thresholds and time resolution after irradiation to a total ionizing dose of 5 kGy and a fluence of  $\Phi_{\mathrm{eq}} = 5\cdot 10^{14}/\mathrm{cm}^2$. The threshold did not change significantly after irradiation, and was measured to be 1963~$\pm$~221~e$^-$ for an unirradiated sensor, and 2035~$\pm$~278~e$^-$ for a sensor irradiated to $\Phi_{\mathrm{eq}}= 5\cdot 10^{14}/\mathrm{cm}^2$ \cite{hiti2021}.

\section{Time resolution of the RD50-MPW2}
With very high rates of collisions expected at the high luminosity LHC and beyond, 4D tracking using time resolution is a requirement for future particle detectors at colliders. The RD50-MPW2 was therefore also characterized in terms of its precision in time.

\begin{figure}[tbh]
    \centering
    \includegraphics[width=0.17\textwidth, trim=2 0 0 3, clip]{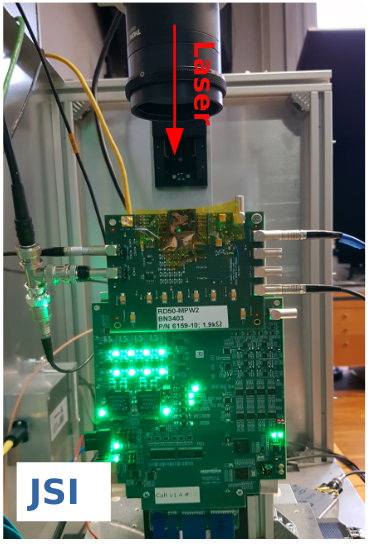}
    \includegraphics[width=0.34\textwidth, trim=2 0 0 3, clip]{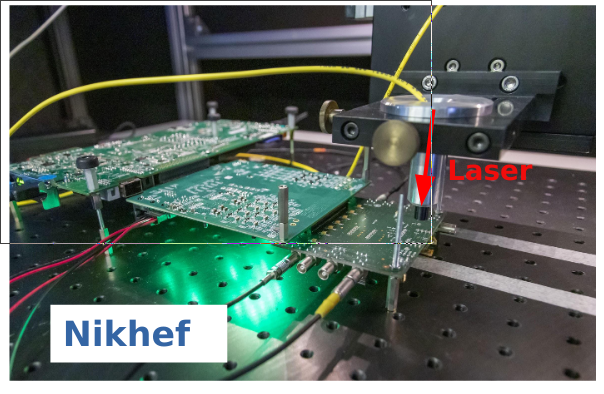}
    \includegraphics[width=0.42\textwidth, trim=2 0 0 3, clip]{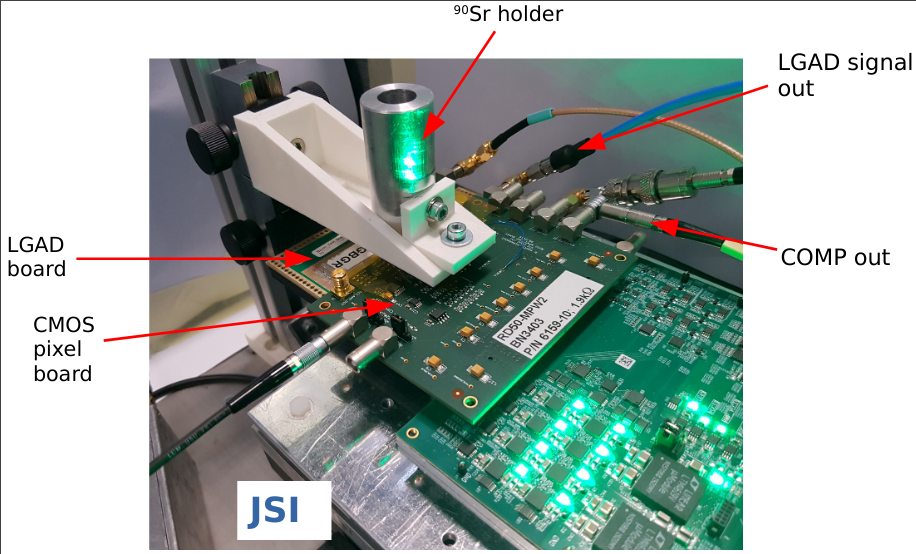}
    \caption{Setup with a 1064 nm infrared laser using edge-TCT at the Jo\v{z}ef Stefan Institute (left) and using back-TCT at Nikhef (middle), and setup with an 18 MBq $^{90}$Sr source at a rate of 1 event per minute at the Jo\v{z}ef Stefan Institute with an $1~\times~1~$mm$^2$ LGAD as reference. Figures from \cite{debevcrd50}.}
    \label{fig:lasersetups}
\end{figure}

\begin{figure}[tbh]
    \centering
    \includegraphics[width=0.42\textwidth, trim=2 0 0 13, clip]{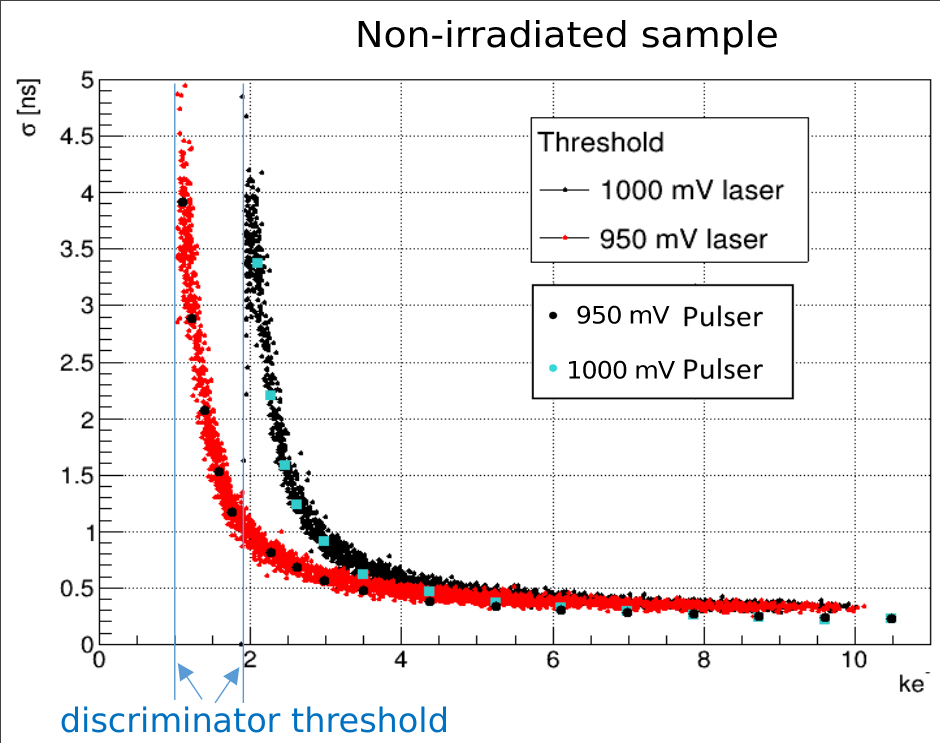}
    \includegraphics[width=0.42\textwidth, trim=2 0 0 13, clip]{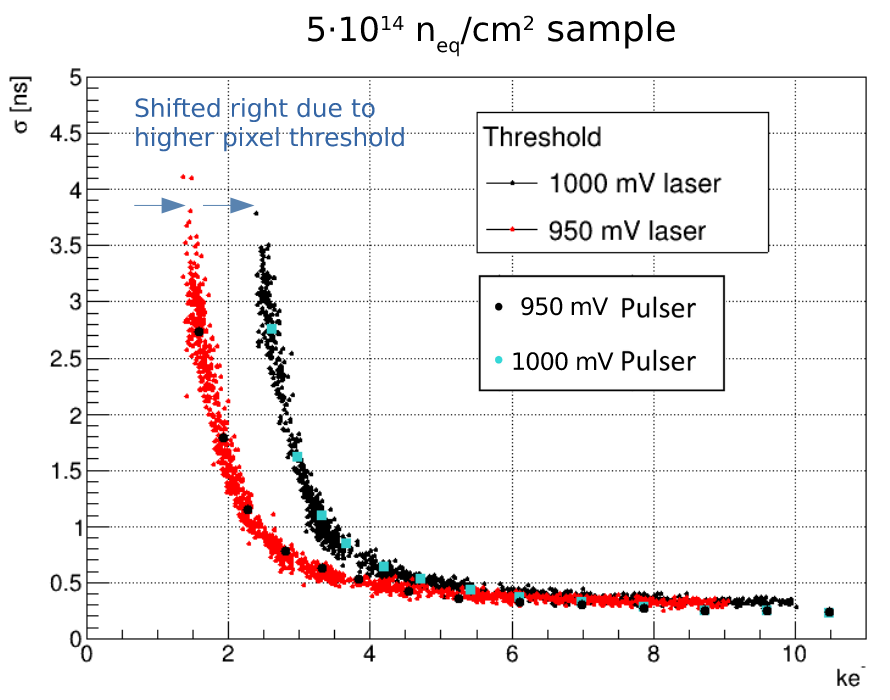}
    \caption{Time resolution from edge-TCT measurements for the RD50-MPW2 for an unirradiated sample (left) and for a sensor irradiated with neutrons to a fluence of $\Phi_{\mathrm{eq}} = 5\cdot     10^{14}/\mathrm{cm}^2$ (right). The laser time resolution with respect to the input pulse is given in the black and red distributions for 1000 mV and 950 mV discriminator thresholds, respectively. The time resolution of the sensor is displayed with blue and black dots in for the same thresholds. Figure from \cite{debevcrd50}.}
    \label{fig:timeresljubljanaedgetct}
\end{figure}

Using a single photon absorption laser and a radioactive source, the time resolution of the second RD50 HV CMOS submission was measured using the spread in the distribution of the time difference between 1) the comparator output and the pulse sent to the laser or 2) the comparator output and the time of arrival with an LGAD, respectively. Pictures of the setups are shown in Fig. \ref{fig:lasersetups}. For a sensor with a resistivity of 1.9 k$\Omega\cdot$cm, 1064 nm infrared laser pulses were sent to a sensor biased at 100 V with two set discriminator thresholds. The time resolution for one continuous pixel is shown for different laser-induced charges in Fig. \ref{fig:timeresljubljanaedgetct}. For a higher threshold, higher input charge is needed for the same time resolution. The time resolution behaves asymptotically towards higher induced charges and reaches a minimum of 300 ps even after irradiation. Note that the depletion depth of the sensor decreases by a large fraction after irradiation \cite{irrad}.

\begin{figure}[tbh]
    \centering
    \includegraphics[width=0.47\textwidth, trim=2 0 0 3, clip]{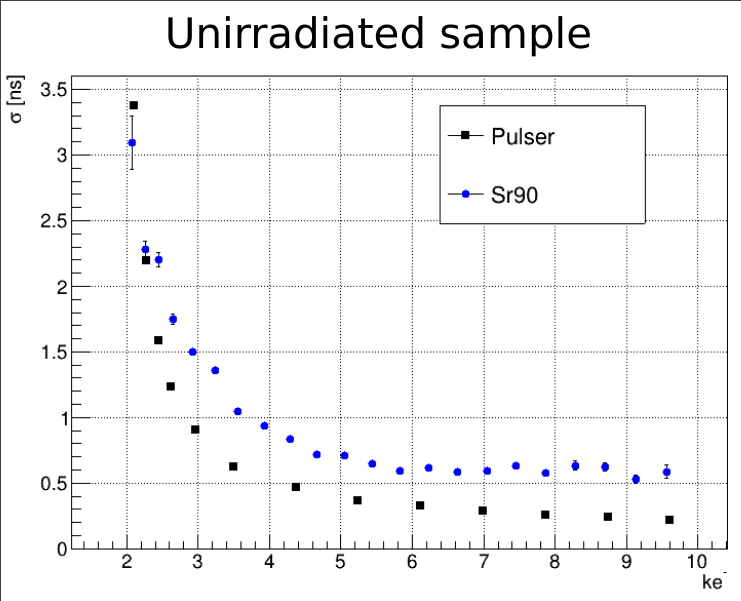}
    \includegraphics[width=0.47\textwidth, trim=2 0 0 3, clip]{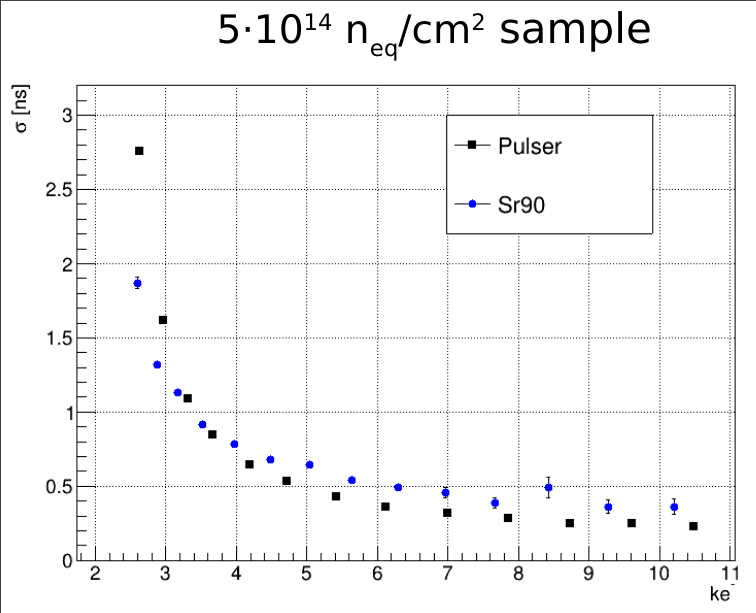}
    \caption{Time resolution from a measurement with a source and LGAD as reference for the RD50-MPW2 for an unirradiated sample (left), and for a sensor irradiated to a fluence of $\Phi_{\mathrm{eq}} = 5\cdot     10^{14}/\mathrm{cm}^2$ (right). The time resolution of the input pulse with respect to the LGAD is given in the black, and the time resolution of the sensor is displayed with blue dots. Figure from \cite{debevcrd50}.}
    \label{fig:timeresljubljanalgad}
\end{figure}

The time resolution was also measured with a 18 MBq $^{90}$Sr source and an LGAD as reference, as shown in Fig. \ref{fig:timeresljubljanalgad}. In this case, the time resolution is worse in the unirradiated sample, where around 600 ps were measured, whereas after irradiation a time resolution of 350 ps only was measured. The worse time resolution before irradiation can be explained by events in which significant charge is collected by diffusion, which is a slow process. After irradiation, the contribution from diffusion to collected charge is much smaller as a result of
trapping and recombination of charges. This is not seen with the laser measurements in Fig. \ref{fig:timeresljubljanaedgetct}, as there the charge is deposited geometrically differently than in case of a minimum ionizing particle. In this case, only signals from the laser in the central part of the depleted region are included and the slower, longer signals are not seen.

\begin{figure}[tbh]
    \centering
    \includegraphics[width=0.47\textwidth, trim=0 0 500 0, clip]{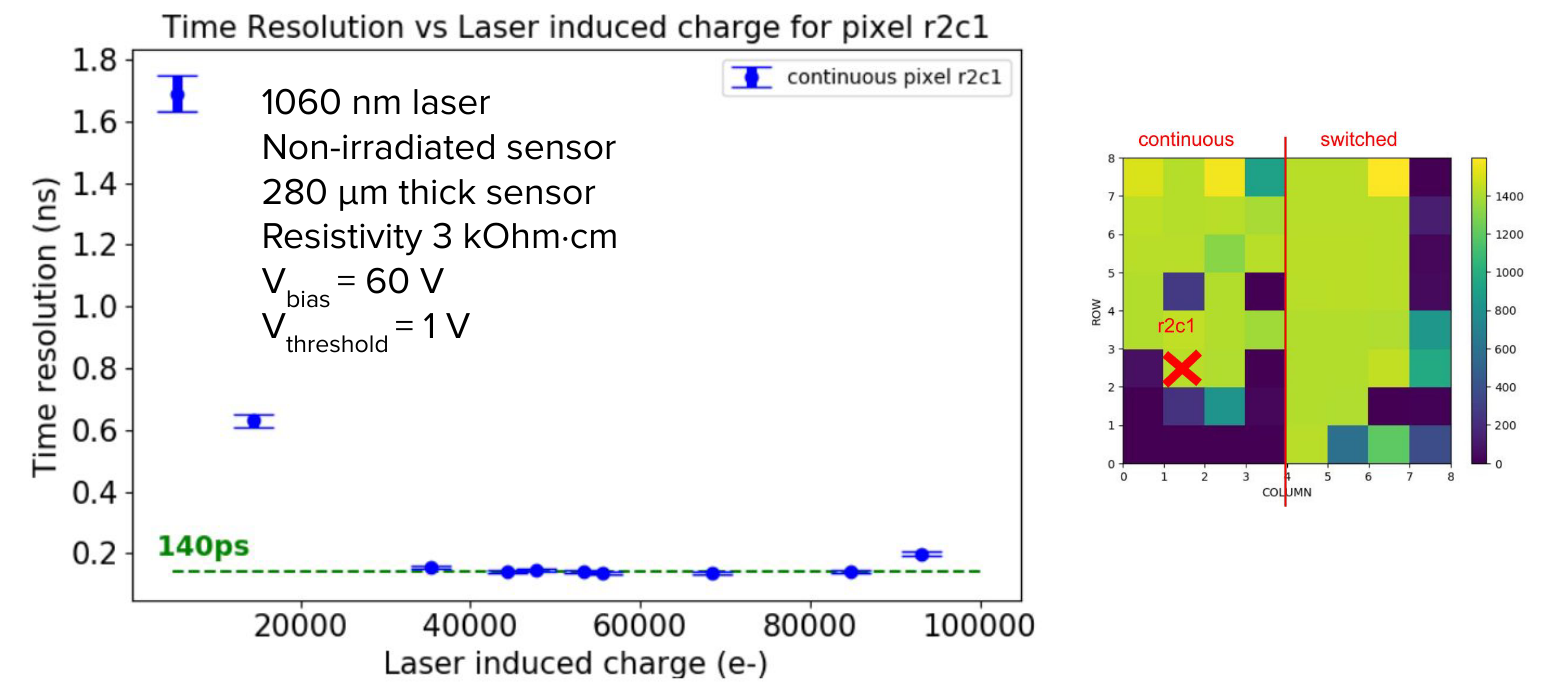}
    \includegraphics[width=0.27\textwidth, trim=1100 100 0 0, clip]{mpw2timeres.png}
    \caption{Time resolution measured with an infrared laser for various laser-induced charges for a single continuous reset pixel. The time resolution reaches 140 ps for larger induced charges. Figure from \cite{christinathesis}.}
    \label{fig:mpw2timeres}
\end{figure}

The time resolution was also measured for a 3~k$\Omega\cdot$cm sample at a bias voltage of 60 V with a threshold of 1000 mV and a 1060 nm infrared laser and a continuous pixel in row 2 column 1, as shown in Fig. \ref{fig:mpw2timeres}. For higher laser-induced charges, the time resolution asymptotically reaches 140 ps; these charges, however, are far beyond those corresponding to a minimum ionizing particle.

\section{Imaging with a two-photon absorption laser}
A two-photon absorption (TPA) laser allows for 3D imaging of semiconductors.
This is one of the best techniques that allows for creating a detailed picture of a sensor, and can be complemented by the proton microprobe technique using real particles \cite{rd50no39}.
An example of the TPA imaging technique used with HV CMOS can be found in \cite{vilatpa}. At the CERN Solid State Detector lab, a 1550 nm two-photon absorption laser was used to create images of the previous RD50 HV CMOS sensor RD50-MPW2. The 400 fs laser has a frequency of 8.2 MHz and an acousto-optic modulator to select a frequency of 1 kHz. Single shots allow for pulses hitting the sensor within its response time. The laser has a variable Z focus and a hexapod for movements in X and Y. The setup is electromagnetic interference isolated. Such a TPA laser is available in 5 high energy physics institutes: CERN (CH), JSI Ljubljana (SI), IFCA (ES), Nikhef (NL), and Lancaster (UK).

\begin{figure}[tbh]
    \centering
    \includegraphics[width=0.42\textwidth, trim=18 2 2 2, clip]{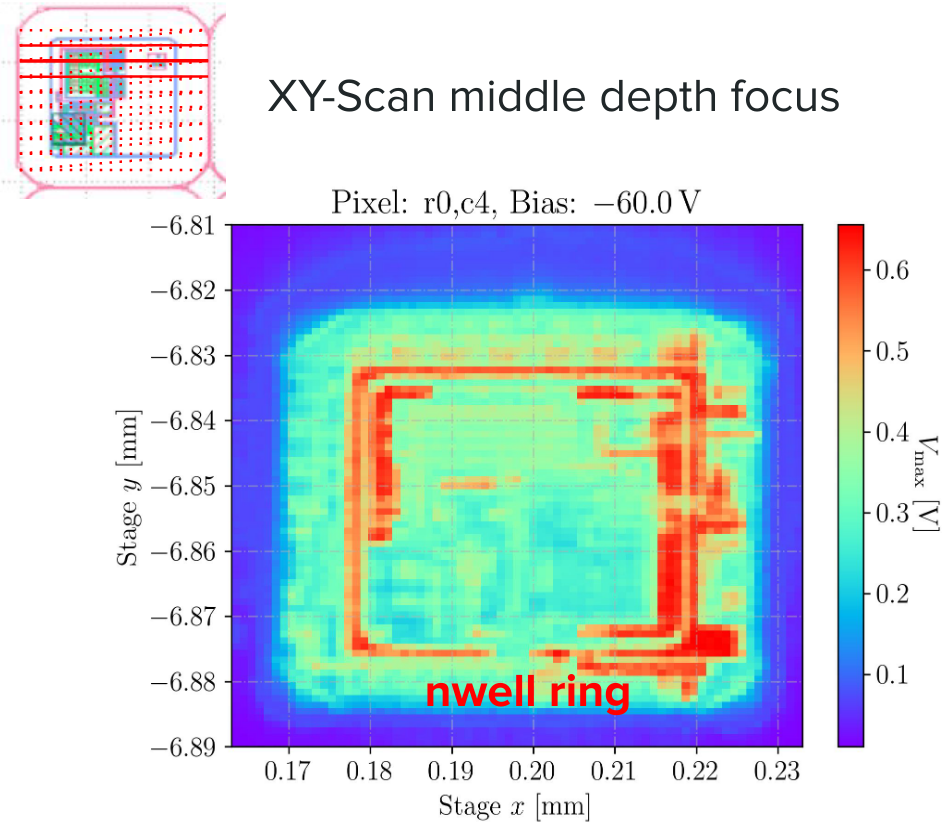}
    \includegraphics[width=0.42\textwidth]{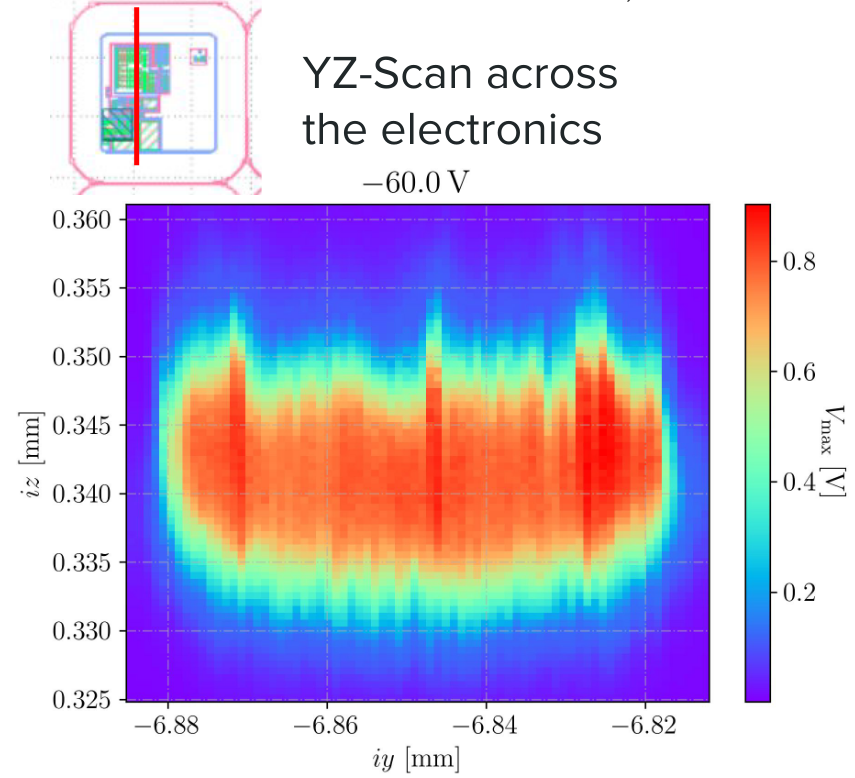}
\caption{Left: image made with a two-photon absorption laser through X and Y at middle depth focus. Back reflection for this focal point near the metallization surface creates a large signal, clearly locating the electronics position in the pixel and confirming the aiming accuracy.
Right: image made with a two-photon absorption laser through Y and Z across the electronics. If the laser hits a depleted region, a large signal is created. This yields a sensitivity map from the voxel position data with the amplitude of the detector signal output. Figures from \cite{tpard50meeting}.}
    \label{fig:tpa}
\end{figure}

A TPA scan of a sensor in X and Y, with Z the depth of the sensor, can map the charge collection efficiency in the pixel volume.
Images made of the RD50-MPW2 using 0.25 nJ per pulse show that the chip electronics that cause back reflections were no obstacle to the mapping and the aiming was accurate, see Fig. \ref{fig:tpa}. In this XY scan through the middle depth focus of the sensor, at $-49 \pm 3~\mu$m, the n-well ring is clearly visible.

The TPA method can also be used to make detailed maps of the electric field. Inhomogeneities in the electric field can affect the depletion region, which can be mapped in detail by scanning a sensor through the depth Z and Y. In Fig. \ref{fig:tpa}, the active depleted region is clearly visible from an YZ-scan across the electronics over the depth of the sensor.

\section{Conclusion and outlook}
The new HV CMOS sensor designed by the RD50 collaboration shows promising results such as a breakdown voltage at 120 V. It has an in-pixel digital readout, advanced periphery that allows for multi-pixel readout. The previous RD50 HV CMOS sensor had a good threshold after irradiation to fluences of  $\Phi_{\mathrm{eq}}~=~5\cdot~10^{14}/\mathrm{cm}^2$ and a time resolution of 300 ps for laser-induced charges comparable to a minimum ionizing particle also after irradiation. Detailed images were created using the two photon absorption laser technique that gave insight into the charge collection efficiency and depletion volume of the previous HV CMOS sensor, the RD50-MPW2. The RD50 CMOS group now plans to further characterize the latest prototype, the RD50-MPW3 in the laboratory and repeat measurements for irradiated sensors. With these results, radiation tolerance will be improved in a next design.

\end{document}